\documentclass[letterpaper, 10 pt, conference]{ieeeconf}  

\IEEEoverridecommandlockouts                              

\usepackage{amsmath}
\usepackage{amssymb}
\usepackage{tikz}
\usepackage{pgfplots}
\usepackage{graphicx}
\usepackage[]{algorithm2e}
\usepackage[caption=false,font=footnotesize]{subfig}
\usepackage{xcolor}

\newcommand{\until}[2]{\, \mathcal{U}_{[{#1},{#2}]} \, }

\newcommand{\rs}[1]{\rho^{#1}(\boldsymbol{x},k)}
\newcommand{\rss}[2]{\rho^{#1}(\boldsymbol{x},{#2})}

\newcommand{\A}[3]{\mathcal{A}^{#1}(\boldsymbol{#2},#3) }

\newcommand{\As}[3]{\mathcal{A_S}^{#1}(\boldsymbol{#2},#3) }
\newcommand{\AAaas}[3]{\mathcal{A_S}^{#1}({\boldsymbol{#2}},#3) }

\newcommand{\At}[1]{c_{temp}^{#1}}

\newcommand{\re}[1]{{\color{black}#1}}
\newcommand{\ree}[1]{{\color{black}#1}}
\newcommand{\defbreak}{\vspace{0.05cm}}

\newtheorem{definition}{Definition} 
\newtheorem{problem}{Problem} 
\newtheorem{theorem}{Theorem} 
\newtheorem{corollary}{Corollary} 
\newtheorem{remark}{Remark} 

\allowdisplaybreaks

\title{\LARGE \bf
Robust Motion Planning employing Signal Temporal Logic*
}

\author{Lars Lindemann$^{1}$ and Dimos V. Dimarogonas$^{1}$
\thanks{*This work was supported in part by the Swedish Research Council (VR), the European Research Council (ERC), the Swedish Foundation for Strategic Research (SSF) and the Knut and Alice Wallenberg Foundation (KAW).}
\thanks{$^{1}$The authors are with the Department of Automatic Control, School of Electrical Engineering, Royal Institute of Technology (KTH), 100 44 Stockholm, Sweden. 
        {\tt\small llindem@kth.se (L. Lindemann), dimos@kth.se (D.V. Dimarogonas)}.}%
}

\begin{document}

\maketitle
\thispagestyle{empty}
\pagestyle{empty}

\begin{abstract}

Motion planning classically concerns the problem of accomplishing a goal configuration while avoiding obstacles. However, the need for more sophisticated motion planning methodologies, \re{taking temporal aspects into account,} has emerged. \re{To address this issue}, temporal logics have recently been used to formulate such advanced specifications. This paper will consider Signal Temporal Logic in combination with Model Predictive Control. A robustness metric, called Discrete Average Space Robustness, is introduced and used to maximize the satisfaction of specifications which results in a natural robustness against noise. The comprised optimization problem is convex and formulated as a Linear Program.

\end{abstract}

\section{Introduction}

A new approach to the motion planning problem has evolved over the past years by formulating \re{system} specifications in temporal logics. Especially linear-time logics have established their application with a focus on Linear Temporal Logic (LTL) as in \cite{gazit1,guo2015multi} and also on Metric Interval Temporal Logic (MITL) \cite{nikou2016cooperative}. LTL and MITL control synthesis uses automata representation of the specification while abstracting the workspace into a transition system. \re{The state space of these automata and their product usually get huge and result in the state explosion problem \cite{baier}}.

Recently, research has been focusing on Signal Temporal Logic (STL), which was introduced in \cite{maler1} within the context of monitoring temporal properties. STL comprises of quantitative time properties and \re{additionally Space Robustness (SR) as introduced in \cite{donze2}, a special case of the robust semantics of \cite{fainekos2009robustness}}. As a consequence it is possible to measure the satisfaction of a specification, i.e., how well a specification is satisfied. \re{Note that the general theoretical concept has already been introduced in \cite{fainekos2009robustness} where it is proven that robust semantics, also denoted as robustness estimates, are an under-approximation of the robustness degree.} This opposes LTL  where only a boolean \re{satisfaction} is given. STL can be used for control synthesis together with Model Predictive Control (MPC) as in our previous work \cite{lindemann_1}. A different MPC approach is used in \cite{raman1} and \cite{sadraddini} where \re{SR} is incorporated into a Mixed Integer Linear Program (MILP). 

Our contribution can be summarized as follows: First, we introduce novel \re{robust semantics}, namely Discrete Average Space Robustness (DASR). DASR has the advantage that it considers average satisfaction, \re{whereas} SR in \cite{donze2} focuses on the worst case scenario, i.e., on a time instant where a specification is least satisfied. In general, \re{it is expected that} DASR  will give better control performance than SR. Second, DASR is directly incorporated into the cost function of a linear MPC framework. Hence, we directly maximize the robustness of satisfying a specification against noise. Third, this new approach is applied to the motion planning problem. \re{The concepts of past satisfaction and recursive feasibility used in this paper are related to those of \cite{sadraddini}. However, our focus is on the robust formulation of linear temporal operators, hence ending up with an efficient encoding as a Linear Program (LP), opposed to the MILP approach in \cite{raman1} and \cite{sadraddini}.} \ree{Therefore, we change and simplify the Space Robustness semantics.}

The remainder of this paper is organized as follows: Section \ref{sec:preliminaries} introduces Signal Temporal Logic, Discrete Average Space Robustness and the problem formulation. Section \ref{sec:our_approach} presents the proposed solution and suggests that the methodology may be suitable for motion planning. The problem solution is verified in section \ref{sec:case_study} by simulations. Conclusion and outlook are provided in section \ref{sec:diskussion_futurework}.

\section{Preliminaries}
\label{sec:preliminaries}

Scalar quantities are denoted as lowercase, non-bold letters $x$. Column vectors are lowercase, bold letters $\boldsymbol{x}$ and matrices are denoted as uppercase, non-bold letters $X$. True and false are denoted by $\top$ and $\bot$; $\otimes$  denotes the Kronecker product while $\boldsymbol{1}_N$ and $\boldsymbol{0}_N$ are vectors containing $N$ ones and zeros, respectively. We denote $E(n,:)$ as the $n$-th row and $E(:,n)$ as the $n$-th column of $E$. \re{Since we deal with discrete-time logics, $[a,b]$ will abbreviate a discretized finite set $\{a,a+1,\hdots,b\}$ where $a$, $b$ with $a\le b$ are integers.}

\subsection{Signals and Systems}
Let $\boldsymbol{x}(k)$, $\boldsymbol{y}(k)$ and $\boldsymbol{u}(k)$ denote the state, output and input, respectively. We consider linear, time-invariant systems in discrete time as
\begin{subequations}\label{system_discrete}
\begin{align}
\boldsymbol{x}(k+1) &= A\boldsymbol{x}(k)+B\boldsymbol{u}(k)\\
\boldsymbol{y}(k)&=\boldsymbol{x}(k),
\end{align}
\end{subequations}
where $A \in \mathbb{R}^{n\times n}$ and $B \in \mathbb{R}^{n\times m}$. 

\subsection{Signal Temporal Logic}
Signal Temporal Logic is a predicate logic based on signals, hence allowing quantitative specifications in space and time. \re{STL consists of predicates $\mu$ that are obtained after evaluation of a function $f(\boldsymbol{x})$ as follows}
\begin{align}
 \mu=
 \begin{cases} 
 \top \text{ if } f(\boldsymbol{x})\ge0\\
 \bot \text{ if } f(\boldsymbol{x})<0.
 \end{cases}
 \end{align}
Hence, $f(\boldsymbol{x})$ determines the truth value of $\mu$ and maps from $\mathbb{R}^n$ to $\mathbb{R}$, whereas $\mu$ maps from $\mathbb{R}^n$ to $\mathbb{B}$;  $\mu$ can be an element of the set $P=\{ \mu_1, \mu_2, \cdots, \mu_{G_\mu} \}$, where $G_\mu$ indicates the number of predicates. Predicates can be expressed as
\begin{align}
\boldsymbol{z}(k) = 
\begin{bmatrix}
f_1(\boldsymbol{x}(k))&
\hdots&
f_{G_\mu}(\boldsymbol{x}(k))
\end{bmatrix}^T=C\boldsymbol{x}(k)+\boldsymbol{c},
\label{eq:stacked_pred}
\end{align}
where $C\in\mathbb{R}^{G_{\mu} \times n}$ and $\boldsymbol{c} \in \mathbb{R}^{G_{\mu}}$ are defined according to the specifications. Note that the mapping in (\ref{eq:stacked_pred}) is affine. In the remainder, single predicates will be abbreviated by 
$z_i(k)=f_i(\boldsymbol{x}(k)) \text{ with } i\in\{1,2,\cdots, G_\mu\}$
for the sake of readability, where the index $k$ might be dropped if it is clear from the context. Inserting the solution $\boldsymbol{x}(k)$ of (\ref{system_discrete}) with initial time $k_0$ into (\ref{eq:stacked_pred}) we can calculate the stacked predicate vector $\boldsymbol{z}_{st}$ for a prediction horizon $N$ as
\begin{align}\label{eq:transformation}
\boldsymbol{z}_{st}=H_1\boldsymbol{x}(k_0)+H_2\boldsymbol{u}_{st}+\boldsymbol{1}_N\otimes \boldsymbol{c},
\end{align}
where
$\re{\boldsymbol{z}_{st}}=\begin{bmatrix}
\boldsymbol{z}(k_0+1)&
\boldsymbol{z}(k_0+2)&
\hdots&
\boldsymbol{z}(k_0+N)
\end{bmatrix} ^T$, 
$\re{\boldsymbol{u}_{st}}=\begin{bmatrix}
\boldsymbol{u}(k_0)&
\boldsymbol{u}(k_0+1)&
\hdots&
\boldsymbol{u}(k_0+N-1)
\end{bmatrix}^T$,
 $\re{H_1}=\begin{bmatrix}
CA&
CA^2&
\hdots&
CA^N
\end{bmatrix}^T$ and 
$\re{H_2}=\begin{bmatrix}
CB 	& 0	& \cdots  	& 0 \\
CAB 	& CB	 & \cdots	& 0 \\
\vdots 	& \vdots & \ddots		& \vdots \\
CA^{N-1}B 	&CA^{N-2}B 	& \cdots 	& CB \\
\end{bmatrix}.$ By using predicates, STL formulas can be assembled. Throughout this paper we will assume that a STL formula is in Positive Normal Form (PNF)\cite{baier}. This means that no negations ($\neg$) occur within a formula except if they are in front of predicates. The STL syntax, given in Backus-Naur form, defines rules to form formulas as
\begin{align}
\phi \; ::= \; \top \; | \; \mu \; | \; \neg \phi \; | \; \phi \wedge \psi \; | \; \phi  \until{a}{b} \psi\;,
\end{align}
where $\mu \in P$ and $\phi$, $\psi$ are STL formulas. The temporal until-operator $\until{a}{b}$ is time bounded with $[a,b]$. Conjunction, eventually-operator and always-operator  can be derived as $\phi\vee\psi=\neg(\neg\phi\wedge\neg\psi)$, $F_{[a,b]}\phi=\top  \until{a}{b} \phi$ and $G_{[a,b]}\phi = \neg F_{[a,b]}\neg\phi$. The semantics of STL are introduced in Definition \ref{def:01} \re{where the satisfaction relation $(\boldsymbol{x},k)\vDash \phi$ denotes if the state sequence $\boldsymbol{x}=\boldsymbol{x}(k)\boldsymbol{x}(k+1)\hdots$ satisfies  $\phi$.}
\defbreak
\begin{definition}[\cite{raman1}] The STL semantics are
\begin{align*}
&(\boldsymbol{x},k) \vDash \mu 				 	&\Leftrightarrow	\;\;\; 	&f(\boldsymbol{x}(k))\ge0\\
&(\boldsymbol{x},k) \vDash \neg\mu 			 	&\Leftrightarrow	\;\;\; 	&\neg((\boldsymbol{x},k) \vDash \mu)\\
&(\boldsymbol{x},k) \vDash \phi \wedge \psi 	 	&\Leftrightarrow \;\;\;	 	&(\boldsymbol{x},k) \vDash \phi \wedge (\boldsymbol{x},k) \vDash \psi\\
&(\boldsymbol{x},k) \vDash \phi \until{a}{b} \psi		&\Leftrightarrow \;\;\;	 	&\exists k_1 \in[k+a,k+b]\text{ s.t. }(\boldsymbol{x},k_1)\vDash \psi \\
&									&					&\wedge \forall k_2\in[k,k_1]\text{,}(\boldsymbol{x},k_2) \vDash \phi\\
&(\boldsymbol{x},k) \vDash F_{[a,b]}\phi		 	&\Leftrightarrow \;\;\;	 	&\exists k_1 \in[k+a,k+b] \text{ s.t. }(\boldsymbol{x},k_1)\vDash \phi\\
&(\boldsymbol{x},k) \vDash G_{[a,b]}\phi		 	&\Leftrightarrow \;\;\;		&\forall k_1 \in[k+a,k+b] \text{,}(\boldsymbol{x},k_1)\vDash \phi
\end{align*}
\label{def:01} 
\end{definition}

The length of a formula $h^\phi$, introduced in \cite{sadraddini}, can be interpreted as the horizon that is needed to calculate the satisfaction of a formula. The recursive definition is
$h^\mu = 0$, 
$h^{\neg \phi} = h^\phi$, $h^{\phi \wedge \psi}=h^{\phi \vee \psi} = \text{max}(h^\phi,h^\psi)$, 
$h^{\phi \until{a}{b} \psi} = b+\text{max}(h^\phi,h^\psi)$, 
$h^{G_{[a,b]}\phi}=h^{F_{[a,b]}\phi}=b+h^\phi$.

\subsection{Average Space Robustness}
\ree{Robust semantics have been introduced to state how well a formula is satisfied. Space Robustness, denoted with $\rs{\phi}$, is such a robustness measure which has been introduced in \cite{donze2}. In the control context it has been applied in \cite{raman1,sadraddini}. For the definition of $\rs{\phi}$ we refer the reader to \cite{donze2}.} Space robustness makes extensive use of \ree{min/max-operations to consider the point of weakest/strongest satisfaction within a signal. We propose a novel robustness measure $\A{\phi}{x}{k}$ in Definition \ref{def:4}, called Discrete Average Space Robustness (DASR), where instead average satisfaction is used, i.e., min-operations as $\underset{k_1\in[k+a,k+b]}{\text{min}}\rss{\phi}{k_1}$  are replaced by an average $\frac{1}{b-a+1}\sum_{k^\prime=k+a}^{k+b}\A{\phi}{x}{k_1}$.}
\defbreak
\begin{definition}{Discrete Average Space Robustness (DASR)}
\begin{align*}
\A{\mu}{x}{k} &= f(\boldsymbol{x}(k))\\
\A{\neg\phi}{x}{k} &= -\A{\phi}{x}{k}\\
\A{\phi \wedge \psi}{x}{k} &= \text{min}(\A{\phi}{x}{k},\A{\psi}{x}{k})\\
\A{\phi \vee \psi}{x}{k} &= \text{max}(\A{\phi}{x}{k},\A{\psi}{x}{k})\\
\A{\phi \until{a}{b}\psi}{x}{k}&=\frac{1}{2}\cdot \Biggl[\underset{k_1\in[k+a,k+b]}{\text{max}} \biggl( \frac{1}{k_1-k+1} \\
			& \hspace{4mm}\cdot \sum_{k^\prime=k}^{k_1}\A{\phi}{x}{k^\prime}  + \A{\psi}{x}{k_1} \biggr)\Biggr]\\
\A{F_{[a,b]}\phi}{x}{k}&=\underset{k_1\in[k+a,k+b]}{\text{max}} \A{\phi}{x}{k_1}\\
\A{G_{[a,b]}\phi}{x}{k}&= \frac{1}{b-a+1}\sum_{k^\prime=k+a}^{k+b} \A{\phi}{x}{k^\prime}
\end{align*}
\label{def:4}
\end{definition}
\ree{By  manually choosing $k_1$ as described in \cite{lindemann_1}}, we can remove the max-operations and define a relaxed version of DASR in Definition \ref{def:5}, called Discrete Simplified Average Space Robustness (DSASR) and denoted by $\As{\phi}{x}{k}$. \ree{As an intuition of the $k_1$ calculation, assume $\phi=F_{[a,b]}z_1$ where we set $k_1=k_0+b$. This results in the highest $\AAaas{\phi}{x}{k_0}$ in most cases since the system has the most time to settle and satisfy $\phi$.} 
\defbreak
\begin{definition}{Discrete Simplified Average Space Robustness (DSASR):}
\begin{align*}
\As{\mu}{x}{k} &= f(\textbf{x}(k))\\
\As{\neg\phi}{x}{k} &= -\As{\phi}{x}{k}\\
\As{\phi \wedge \psi}{x}{k} &= \text{min}(\As{\phi}{x}{k},\As{\psi}{x}{k})\\
\As{\phi \vee \psi}{x}{k} &= \text{max}(\As{\phi}{x}{k},\As{\psi}{x}{k})\\
\As{\phi \until{a}{b}\psi}{x}{k}&=\frac{1}{2}\cdot \biggl[ \frac{1}{k_1-k+1} \sum_{k^\prime=k}^{k_1}\As{\phi}{x}{k^\prime} \\
					& \hspace{4mm}+ \As{\psi}{x}{k_1}\biggr]\\
\As{F_{[a,b]}\phi}{x}{k}&= \As{\phi}{x}{k_1}\\
\As{G_{[a,b]}\phi}{x}{k}&= \frac{1}{b-a+1}\sum_{k^\prime=k+a}^{k+b} \As{\phi}{x}{k^\prime}
\end{align*}
\label{def:5}
\end{definition}
\re{Note that the robust semantics in \cite{fainekos2009robustness} and hence also $\rs{\phi}$ from \cite{donze2} are an under-approximation of the robustness degree in \cite{fainekos2009robustness}. However, DASR and DSASR are not such an under-approximation. \ree{This can be seen by considering $\phi=G_{[a,b]}(x>0)$, where it is possible that if $\As{\phi}{x}{0}=\frac{1}{b-a+1}\sum_{k^\prime=a}^{b}x(k^\prime)>0$, there might be a $k_1 \in[a,b]$ s.t. $x(k_1)<0$ and hence $\rss{\phi}{0}=\underset{k_1\in[a,b]}{\text{min}}x(k_1)<0$.}  Subsequently, $\As{\phi}{x}{k}>0 \nRightarrow (\boldsymbol{x},k) \vDash \phi $, whereas $\rs{\phi} > 0 \Rightarrow (\boldsymbol{x},k) \vDash \phi$. However, in this paper additional constraints imposed on the optimization problem will ensure this property. We remark that averaged STL (AvSTL) introduced in \cite{akazaki2015time} is different compared with DASR and DSASR. The averaged temporal operators of AvSTL form a weighted time average over $\rs{\phi \until{a}{b} \psi}$, $\rs{F_{[a,b]} \phi}$ and $\rs{G_{[a,b]} \phi}$, hence not removing min/max-operations and keeping a nonlinear description. This results in nonconvex temporal operators which cause computational burdens in optimization problems. Furthermore, AvSTL considers time and space robustness, while DASR and DSASR only consider the latter. Time robustness is a useful measure yet with the drawback of more complex definitions that can be handled in a monitoring, but hardly within a control context.}

\subsection{Problem Statement}
\label{sec:problem_statement}

\ree{This paper considers a subset of STL, namely 
$\psi_1=z_1 \until{a}{b} z_2$, 
$\psi_2=F_{[a,b]}z_1$, 
$\psi_3=G_{[a,b]}z_1$, 
$\psi_4=\psi_{i_1} \wedge \psi_{i_2} \wedge \cdots \wedge \psi_{i_n}$, 
$\psi_5=\psi_{i_1} \vee \psi_{i_2}  \vee \cdots \vee \psi_{i_n}$ and
$\psi_6=\psi_{i_1} (\vee \text{ or } \wedge) \psi_{i_2}  (\vee \text{ or } \wedge) \cdots (\vee \text{ or } \wedge) \psi_{i_n} \text{ for }i_1,\cdots,i_n\in\{1,2,3\}$. We distinguish between two types of formulas, namely all-time satisfying and one-time satisfying formulas. The former means that the formula $\psi_i$ with $i=1,\hdots,6$ is imposed at every sampling step, i.e., $\phi_i=G_{[0,\infty]}\psi_i$. One-time satisfying formulas are characterized by satisfying the formula once which is denoted by $\phi_7=event  \implies \psi_i$. The boolean variable $event$ is an indicator for the time when $\psi_i$ is triggered.} We will include a notion of past satisfaction in the same vein as in \cite{sadraddini}.  By respecting the prediction horizon $N$ and the formula length $h^{\psi_i}$, we set $k_l=k_0-h^{\psi_i}$ and $k_h=k_0+N-h^{\psi_i}$. The formula length $h^\psi_i$ plays the role of determining how many predicates of the past and the future need to be used given $N$.

\defbreak
\begin{problem}
Given a linear, time-invariant system (\ref{system_discrete}), a STL formula $\phi_i=G_{[0,\infty]}\psi_i$ with $i=1,\hdots,6$, an initial state $\boldsymbol{x}(k_0)$ and a prediction horizon $N\ge h^{\psi_i}$, compute
\begin{subequations}\label{eq:problem1}
\begin{align}
&\underset{\boldsymbol{u}_{st}}{\operatorname{argmax}} \sum_{k^\prime=k_l}^{k_h} \As{\psi_i}{x}{k^\prime} \label{eq:cost_111}\\
\text{s.t. } &\boldsymbol{x}(k+1) = A\boldsymbol{x}(k)+B\boldsymbol{u}(k) \label{eq:constraint_x}\\ 
&(\boldsymbol{x},k) \vDash \psi_i \re{\text{ for all } k\in[k_l,k_h]}.
\end{align}
\end{subequations}
Note that $\boldsymbol{u}_{st}$ directly determines $\boldsymbol{x}$ and consequently shapes $\AAaas{\psi_i}{x}{k}$. Furthermore, solve the same problem for formulas $\phi_7=event  \implies \psi_i$. \ree{In the remainder we will not explicitly mention constraint \eqref{eq:constraint_x} due to space limitations.}
\label{prob:gen}
\end{problem}

\section{Control Strategy}
\label{sec:our_approach}
\re{DSASR is linear and convex in the temporal operators and will hence be included in the cost function of the MPC framework.} We start by investigating the basic temporal operators, i.e., \ree{$\phi_1=G_{[0,\infty]}\big(z_1 \until{a}{b} z_2\big)$, $\phi_2=G_{[0,\infty]}\big(F_{[a,b]}z_1\big)$ and $\phi_3=G_{[0,\infty]}\big(G_{[a,b]}z_1)$.}
\defbreak
\begin{theorem}\label{theorem:1}
\re{The optimization problem \eqref{eq:problem1} subject to the formulas $\phi_1$, $\phi_2$ and $\phi_3$ can be written as a Linear Program.}

\begin{proof}
The operator $\phi_1$ can be formulated as
\begin{subequations}\label{opt_problem}
\begin{align}
\begin{split}
&\underset{\boldsymbol{u}_{st}}{\operatorname{argmax}} \; \frac{1}{2}\cdot \sum_{k^\prime=k_l}^{k_h} \bigg[ z_2\Big(k_1(k^\prime)\Big) \label{eq:un1} \; \\ 
&\hspace{16mm} +\frac{1}{k_1(k^\prime)-k^\prime+1} \sum_{k^{\prime\prime}=k^\prime}^{k_1(k^\prime)}z_1(k^{\prime\prime}) \bigg] \\
\end{split}\\
\text{s.t. } &z_1(k)\ge 0 \; \; \;  \re{\forall \ k^\prime \in [k_l,k_h]}, \forall  k \in [\re{k^\prime},k_1(k^\prime)]\label{eq:1111a}\\
		&z_2\Big(k_1(k^\prime)\Big)\ge 0 \;\;\; \re{\forall \ k^\prime \in [k_l,k_h]}\label{eq:1111b},
\end{align}
\end{subequations}
\re{where an intuition of $k_1(k^\prime)$ has already been given}. The cost function (\ref{eq:un1}) can be reduced to a Linear Program as
\begin{align}
&\underset{\boldsymbol{u}_{st}}{\operatorname{argmax}} \;  \boldsymbol{1}_N^T E\; \boldsymbol{z}_{all} \label{eq:lin_prog},
\end{align}
where $\boldsymbol{z}_{all}$ concatenates the first $G_\mu(N-h^{\psi_1})$ elements of $\boldsymbol{z}_{st}$ with past predicates from time $k_l$ as follows:
\begin{align}
\boldsymbol{z}_{all}=\begin{bmatrix}
\boldsymbol{z}(k_l)&
\hdots&
\boldsymbol{z}(k_0)&
\boldsymbol{z}_{st}(1:G_\mu(N-h^{\psi_1}))
\end{bmatrix}^T
\label{eq:z_all}
\end{align}
The $E$ matrix depends on $k_1(k^\prime)$ and is of size $E\in\mathbb{R}^{N\times (G_\mu N)}$. Note that the columns of $E$ are associated (multiplied) with $\boldsymbol{z}_{all}$ and depend on the number of predicates $G_\mu$ and the prediction horizon $N$. In other words, $E \boldsymbol{z}_{all}$ is a vector of size $N$ that consists of each sum element in (\ref{eq:un1}), i.e.,
$z_2\Big(k_1(k^\prime)\Big) +\frac{1}{k_1(k^\prime)-k^\prime+1} \sum_{k^{\prime\prime}=k^\prime}^{k_1(k^\prime)}z_1(k^{\prime\prime})$ for $k^\prime\in\{k_l,\hdots,k_h\}$.  Each row of $E$ is associated with $k^\prime$ and hence represents a different time instant in the sum $\sum_{k^\prime=k_l}^{k_h}$ of (\ref{eq:un1}). Consequently, multiplying $\boldsymbol{1}_N^T$ with $E \boldsymbol{z}_{all}$ amounts to the complete cost function given in (\ref{eq:un1}). For the until-operator, where $G_\mu=2$, $E$ can be formed using the following step-by-step procedure:
\begin{enumerate}
\item Start with $k^\prime=k_l$ and set $i=1$.
\item Form a row vector of size $2N$, where the $\left(2(h^{\psi_1}+k_1(k^\prime))\right)$-th column is set to $\frac{1}{2}$, which corresponds to the term $\frac{1}{2} z_2\Big(k_1(k^\prime)\Big)$. Set all odd columns between the columns $2(i-1)+1$ and $2(h^{\psi_1}+k_1(k^\prime))$ to $ \frac{1}{2(k_1(k^\prime)-k^\prime+1)}$, which corresponds to the term $ \frac{1}{2(k_1(k^\prime)-k^\prime+1)} \sum_{k^{\prime\prime}=k^\prime}^{k_1(k^\prime)}z_1(k^{\prime\prime})$. Set all other elements to $0$ and make this row vector the $E$ matrix if $i=1$. Otherwise, append this row vector to $E$.
\item Stop if $k^\prime=k_h$, else increase $k^\prime$ and $i$ by $1$ and go back to step 2).
\end{enumerate}

The operator $\phi_2$ can be formulated as
\begin{subequations}\label{eq:opt_ev}
\begin{align}
&\underset{\boldsymbol{u}_{st}}{\operatorname{argmax}} \sum_{k^\prime=k_l}^{k_h} z_1\Big(k_1(k^\prime)\Big)\\
\text{s.t. } 	&z_1\Big(k_1(k^\prime)\Big)\ge 0 \;\;\; \re{\forall \ k^\prime \in [k_l,k_h]} \label{eq:1111c}
\end{align}
\end{subequations}
and reduced to the cost function as in \eqref{eq:lin_prog} with $G_\mu=1$ and $E$ formed according to the following procedure:
\begin{enumerate}
\item Start with $k^\prime=k_l$ and $i=1$.
\item Form a row vector of size $N$, where the $\left(h^{\psi_2}+k_1(k^\prime)\right)$-th column is set to $1$. Set all other elements to $0$ and make this row vector the $E$ matrix if $i=1$. Otherwise, append this row vector to $E$.
\item Stop if $k^\prime=k_h$, else increase $k^\prime$ and $i$ by $1$ and go back to step 2).
\end{enumerate}

The operator $\phi_3$ can be formulated as
\begin{subequations}
\begin{align}
&\underset{\boldsymbol{u}_{st}}{\operatorname{argmax}} \sum_{k^\prime=k_l}^{k_h} \frac{1}{b-a+1}\sum_{k^{\prime\prime}=k^\prime+a}^{k^\prime+b} z_1(k^{\prime\prime}) \\
\text{s.t. } 	&z_1(k)\ge 0  \; \; \; \re{\forall \ k^\prime \in [k_l,k_h]}, \forall \; k \in [k^\prime+a,k^\prime+b]\label{eq:1111d}
\end{align}
\end{subequations}
and reduced to the cost function as in \eqref{eq:lin_prog} with $G_\mu=1$ and $E$ being formed as follows:
\begin{enumerate}
\item Start with $k^\prime=k_l$ and $i=1$.
\item Form a row vector of size $N$, where all columns in $[a+i,b+i]$ are set to $\frac{1}{b-a+1}$. Set all other elements to $0$ and make this row vector the $E$ matrix if $i=1$. Otherwise, append this row vector to $E$.
\item Stop if $k^\prime=k_h$, else increase $k^\prime$ and $i$ by $1$ and go back to step 2).
\end{enumerate}

\end{proof}
\end{theorem}

To illustrate how $E$ looks like, assume that $N=4$, $h^{\psi_1}=2$ and $k_0=0$. Also assume that
$
k_1(k^\prime)=
\begin{cases}
0 \text{ if $k^\prime = \{-1,0$\}}\\
2 \text{ if $k^\prime = \{1,2$\}}
\end{cases}
$.
For the until-operator, we get
$
E=\frac{1}{2}\cdot
\begin{bmatrix}
\frac{1}{2} & 0 & \frac{1}{2} & 1 & 0 & 0 & 0 & 0\\
0 & 0 & 1 & 1 & 0 & 0 & 0 & 0\\
0& 0 & 0 & 0 & \frac{1}{2} & 0 & \frac{1}{2} & 1\\
0 & 0 & 0 & 0 & 0& 0 & 1 & 1
\end{bmatrix},
$
where the rows $E(1,:)$ and $E(2,:)$ represent $\AAaas{\psi_1}{x}{k^\prime}$ at times $k^\prime=-1$ and $k^\prime=0$. The columns $E(:,1)$, $E(:,3)$ and $E(:,5)$ are associated with $z_1(-1)$, $z_1(0)$ and $z_1(1)$, whereas $E(:,2)$, $E(:,4)$ and $E(:,6)$ are associated with $z_2(-1)$, $z_2(0)$ and $z_2(1)$. Recall that $\boldsymbol{z}_{all}=\begin{bmatrix}
z_1(-1) & z_2(-1) & z_1(0) & z_2(0) & \hdots & z_1(2) & z_2(2)
\end{bmatrix}^T$.

Next, we will investigate conjunctions of the form $\phi_4$ as introduced in section \ref{sec:problem_statement}. 
\defbreak
\begin{theorem}\label{theorem:2}
\re{The optimization problem \eqref{eq:problem1} subject to the formula $\phi_4$ can be written as a Linear Program.}

\begin{proof}
First, we assume $\psi_4 = \psi_i \wedge \psi_j$ where $i,j\in\{1,2,3\}$. The problem can be expressed as
\begin{subequations}\label{eq:conj_opt}
\begin{align}
&\underset{\boldsymbol{u}_{st}}{\operatorname{argmax}} \sum_{k^\prime=k_l}^{k_h} \text{min}(\As{\psi_i}{x}{k^\prime},\As{\psi_j}{x}{k^\prime})\label{eq:conj_opt111}\\
\text{s.t. } &\re{\At{\psi_i}} \text{ and } \re{\At{\psi_j}} ,
\end{align}
\end{subequations}
\ree{where $\At{\psi_i}$ is a shortcut for the constraints \eqref{eq:1111a} and \eqref{eq:1111b} if $i=1$, \eqref{eq:1111c} if $i=2$ and \eqref{eq:1111d} if $i=3$, i.e, $\At{\psi_1}:=$\eqref{eq:1111a}$\wedge$\eqref{eq:1111b}, $\At{\psi_2}:=$\eqref{eq:1111c} and $\At{\psi_3}:=$\eqref{eq:1111d}.} Note that the cost function in (\ref{eq:conj_opt111}) is a sum of finite elements, which can be written as
$
\text{min}(\As{\psi_i}{x}{k_l},\As{\psi_j}{x}{k_l}) 
+\text{min}(\As{\psi_i}{x}{k_h},\As{\psi_j}{x}{k_h}).
$
Since (\ref{eq:conj_opt111}) is a max-min problem, the expression can be simplified by introducing an additional decision variable $u_{x,n}$ with $n\in\{1,\hdots,N\}$ for each sum element. First, define the vector 
$\boldsymbol{u}_x=\begin{bmatrix}
u_{x,1} &
\hdots&
u_{x,N}&
\boldsymbol{u}(k_0)&
\hdots&
\boldsymbol{u}(k_0+N-1)
\end{bmatrix}^T
$
and rewrite problem (\ref{eq:conj_opt}) as
\begin{subequations}\label{eq:conj_ref}
\begin{align}
&\underset{\boldsymbol{u}_x}{\operatorname{argmax}} \; \sum_{i=1}^{N}u_{x,i} \\
\text{s.t. } &u_{x,1} \le  \As{\psi_i}{x}{k_l} \label{eq:conj_ref111a}\\
&u_{x,1} \le  \As{\psi_j}{x}{k_l}\\
& \vdots\\
&u_{x,N} \le  \As{\psi_i}{x}{k_h}\\
&u_{x,N} \le  \As{\psi_j}{x}{k_h} \label{eq:conj_ref111b}\\
&\re{\At{\psi_i}} \text{ and } \re{\At{\psi_j}}.
\end{align}
\end{subequations}
Note that (\ref{eq:conj_ref}) and (\ref{eq:conj_opt}) are equivalent (see \cite{boyd} for similar examples). This is again a Linear Program
$
\underset{\boldsymbol{u}_x}{\operatorname{argmax}}\; \boldsymbol{f}^T\boldsymbol{u}_x,
$
where $\boldsymbol{f}=
\begin{bmatrix}
\boldsymbol{1}_N &
\boldsymbol{0}_{N m}
\end{bmatrix}^T
$. By defining
$
H_{2,man} = \begin{bmatrix} 
\underline{0}_{G_\mu N,N} & H_2
\end{bmatrix}
$
and $\underline{0}_{G_\mu N,N}$ as a matrix consisting of zeros with $G_\mu N$ rows and $N$ columns, the stacked predicate vector from (\ref{eq:transformation}) can be reformulated as
$
\boldsymbol{z}_{st}=H_1\boldsymbol{x}(k_0)+H_{2,man}\boldsymbol{u}_x+\boldsymbol{1}_N\otimes \boldsymbol{c}.
$
Define again $\boldsymbol{z}_{all}$ as in (\ref{eq:z_all}) and reformulate the linear inequalities of (\ref{eq:conj_ref111a}) - (\ref{eq:conj_ref111b}) as
$
Q\boldsymbol{u}_x \le R\boldsymbol{z}_{all}.
$
The $Q$ matrix is given by
$
Q=
\begin{bmatrix}
1 & 0 & \cdots & 0 & 0 & \cdots & 0\\
1 & 0 & \cdots & 0 & 0 & \cdots & 0\\
0 & 1 & \cdots & 0 & 0 & \cdots & 0\\
0 & 1 & \cdots & 0 & 0 & \cdots & 0\\
\vdots & \vdots & \vdots & \vdots & \vdots & \vdots & \vdots 
\end{bmatrix},
$
whereas $R$ depends on the structure of the temporal operators as explained in the sequel. At first, the $E$ matrices, denoted as $E_{\phi_i,r}$ and $E_{\phi_j,r}$, need to be created according to the given rules for $\phi_1$, $\phi_2$ and $\phi_3$. However, for a conjunction more predicates are used than in the case of one temporal operator. Hence, these matrices need to be changed slightly, i.e., for each additional predicate, columns consisting of zeros need to be inserted. For instance, consider $F_{[a,b]}z_i \wedge F_{[a,b]}z_j$ with the corresponding matrices $E_{\phi_{i},r}$ and $E_{\phi_{j},r}$ with $p=N G_\mu$ columns each. Then the matrices
$E_{\phi_{i}} = \begin{bmatrix} E_{\phi_{i},r}(:,1) & \boldsymbol{0}_N & E_{\phi_{i},r}(:,2) & \hdots  & E_{\phi_{i},r}(:,N)  &\boldsymbol{0}_N \end{bmatrix}$ and
$E_{\phi_{j}} = \begin{bmatrix}  \boldsymbol{0}_N & E_{\phi_{j},r}(:,1) & \boldsymbol{0}_N & \hdots & \boldsymbol{0}_N & E_{\phi_{j},r}(:,N) \end{bmatrix}$ have $2p$ columns. This is due to the fact that we now have twice the amount of predicates. Finally, $R$ can be composed as
$
R =
\begin{bmatrix}
E_{\phi_i}(1,:)&
E_{\phi_j}(1,:)&
\hdots&
E_{\phi_i}(N,:)&
E_{\phi_j}(N,:)&
\end{bmatrix}^T.
$

An extension to more than one operator ($\psi_i \wedge \psi_j \wedge \psi_k \wedge \cdots$) can easily be handled by adding one additional constraint for each added conjunction. For instance, for $\psi_i \wedge \psi_j \wedge \psi_k$ the constraints 
$
u_{x,1} \le  \As{\psi_k}{x}{k_l}$, $\hdots$, 
$u_{x,N} \le  \As{\psi_k}{x}{k_h}$, 
$\re{\At{\psi_k}} $
need to be added to (\ref{eq:conj_ref}).
\end{proof}
\end{theorem}

Disjunction formulas as $\phi_5$ can be handled as follows.
\defbreak
\begin{theorem}\label{theorem:3}
\re{The optimization problem \eqref{eq:problem1} subject to the formulas $\phi_5$ and $\phi_6$ can be written as a Linear Program.}

\begin{proof} 
Again, think of two temporal operators connected by a disjunction $\psi_5 = \psi_i \vee \psi_j$ where $i,j\in\{1,2,3\}$. To approach this problem, calculate the optimal solution and the corresponding optimal input sequence $\boldsymbol{u}_{st,1}^*$ for 
$
\underset{\boldsymbol{u}_{st,1}}{\operatorname{argmax}} \; \sum_{k^\prime=k_0-h^{\psi_i}+1}^{k_0+N-h^{\psi_i}} \As{\psi_i}{x}{k^\prime}$ s.t.
$\At{\psi_i}$ and $\boldsymbol{u}_{st,2}^*$ for
$
\underset{\boldsymbol{u}_{st,2}}{\operatorname{argmax}} \; \sum_{k^\prime=k_0-h^{\psi_j}+1}^{k_0+N-h^{\psi_j}} \As{\psi_j}{x}{k^\prime}$ s.t. $\At{\psi_j}$. The optimal state trajectories that result from the optimal inputs $\boldsymbol{u}_{st,1}^*$ and $\boldsymbol{u}_{st,2}^*$ are denoted by $\boldsymbol{x}_1^*$ and $\boldsymbol{x}_2^*$, respectively. Next, calculate the optimal costs given by
$
C_1 = \sum_{k^\prime=k_0-h^{\psi_i}+1}^{k_0+N-h^{\psi_i}} \AAaas{\psi_i}{x_1^*} {k^\prime}$ and
$C_2 = \sum_{k^\prime=k_0-h^{\psi_j}+1}^{k_0+N-h^{\psi_j}} \AAaas{\psi_j}{x_2^*} {k^\prime}$.
The input corresponding to the biggest $C_i$ will be applied to the system. This procedure can be applied in exactly the same way to solve formulas like $\phi_6$, where additional conjunctions lead to additional $C_i$'s.
\end{proof}
\end{theorem}

Motion planing tasks can be formulated as one-time satisfying STL formulas which are a subclass of all-time satisfying formulas presented so far. Hence, the same methodology can be used in a simplified manner. The sum in the cost function (\ref{eq:cost_111}) reduces to one element 
$
\underset{\boldsymbol{u}_{st}}{\operatorname{argmax}} \; \AAaas{\psi_i}{x}{k_0}.
$
Subsequently, the $E$ matrices derived before simplify to a row vector (recall that each row is associated with $k^\prime$). To illustrate this, consider  $\phi_7=event  \implies \psi_i$ with $i \in \{1,2,3\}$. First, let $E_{\phi_i,one}$ denote the $E$ matrix constructed according to the construction rules for $\phi_i=G_{[0,\infty]}\psi_i$. Let $k_{event}$ indicate for how long $event$ has been activated. Then $E_{\phi_i}$ is constructed by selecting $E_{\phi_i}=E_{\phi_i,one}(h^{\psi_i}-k_{event},:)$, i.e., the ($h^{\psi_i}-k_{event}$)-th row of $E_{\phi_i,one}$. The procedure for formulas $event  \implies \psi_j$ with $j \in \{4,5,6\}$ can mutatis mutandis be adopted. Also recall that $p \implies q$ is equivalent to $\neg p \vee q$.
\begin{remark}
\ree{Theorems \ref{theorem:1}, \ref{theorem:2} and \ref{theorem:3} guarantee that if the optimization problem is feasible it follows $(\boldsymbol{x},k) \vDash \phi$.}
\end{remark}

\re{Finally, we provide a statement about recursive feasibility.
\begin{corollary}
The optimization problem \eqref{eq:problem1} subject to the formulas $\phi_1$ to $\phi_7$ in PNF can be modified such that in case of infeasibility the least violating solution is found.

\begin{proof}
The idea is similar to \cite{sadraddini} and makes use of a slack variable $\xi\ge 0$ . The cost function \eqref{eq:cost_111} is extended to $\sum_{k^\prime=k_l}^{k_h} \As{\psi_i}{x}{k^\prime}-M\xi$, where $M$ is a sufficiently large real number. Next, the constraints in \eqref{eq:1111a}, \eqref{eq:1111b}, \eqref{eq:1111c} and \eqref{eq:1111d} need to be modified to 
$z_1(k) + \xi \ge 0 $, 
$z_2\Big(k_1(k^\prime)\Big)+ \xi \ge 0$, 
$z_1\Big(k_1(k^\prime)\Big) + \xi\ge 0$ and
$z_1(k) +\xi \ge 0 $, respectively. \ree{In contrast to our approach, \cite{sadraddini} includes the slack variable $\xi$ in the predicates as $\boldsymbol{z}_{soft}=\boldsymbol{z}+\boldsymbol{1}_{G_\mu}\xi$. We avoid this since $\boldsymbol{z}$ is part of the cost function that we do not want to alter}.
\end{proof}
\end{corollary}}

\section{Case Study}
\label{sec:case_study}
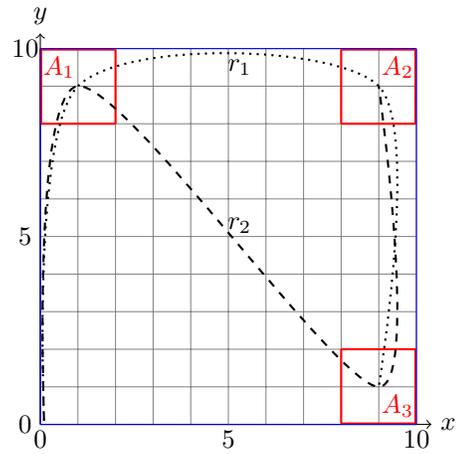
\begin{figure}
\centering
\begin{tikzpicture}[scale=0.5]
    \draw[very thin,color=gray] (0,0) grid (10,10);
    \draw[->,thin] (0,0) -- (10.4,0) node[right] {$x$};
    \draw[->,thin] (0,0) -- (0,10.4) node[above] {$y$};
    \coordinate (p1) at (0.025,0.025);
    \coordinate (p2) at (0.025,9.975);
    \coordinate (p3) at (9.975,0.025);
    \coordinate (p4) at (9.975,9.975);
    \coordinate (p5) at (2,0.025);
    \coordinate (p6) at (2,2);
    \coordinate (p7) at (0.025,2);
    \coordinate (p8) at (8,0.025);
    \coordinate (p9) at (8,2);
    \coordinate (p10) at (9.975,2);
    \coordinate (p11) at (0.025,8);
    \coordinate (p12) at (2,8);
    \coordinate (p13) at (2,9.975);
    \coordinate (p14) at (8,8);
    \coordinate (p15) at (8,9.975);
    \coordinate (p16) at (9.975,8);
    \coordinate (b1) at (0,0);
    \coordinate (b2) at (0,10);
    \coordinate (b3) at (10,0);
    \coordinate (b4) at (10,10);
    \draw[-,color=red,thick] (p2) to (p11);
    \draw[-,color=red,thick] (p2) to (p13);
    \draw[-,color=red,thick] (p13) to (p12);
    \draw[-,color=red,thick] (p11) to (p12);
    \draw[-,color=red,thick] (p4) to (p15);
    \draw[-,color=red,thick] (p4) to (p16);
    \draw[-,color=red,thick] (p15) to (p14);
    \draw[-,color=red,thick] (p16) to (p14);
    \draw[-,color=red,thick] (p3) to (p10);
    \draw[-,color=red,thick] (p3) to (p8);
    \draw[-,color=red,thick] (p10) to (p9);
    \draw[-,color=red,thick] (p8) to (p9);
	\draw[-,color=blue] (b1) to (b2) ;
	\draw[-,color=blue] (b1) to (b3) ;
	\draw[-,color=blue] (b3) to (b4) ;
	\draw[-,color=blue] (b2) to (b4) ;
	\node [red,thick] at (0.5,9.5) {$A_1$};
	\node [red,thick] at (9.5,0.5) {$A_3$};
	\node [red,thick] at (9.5,9.5) {$A_2$};
	\node [thick] at (5.3,9.5) {$r_1$};
	\node [thick] at (5.3,5.3) {$r_2$};
	\node at (-0.4,0) {$0$};
	\node at (0,-0.4) {$0$};
	\node at (-0.4,5) {$5$};
	\node at (5,-0.4) {$5$};
	\node at (-0.4,10) {$10$};
	\node at (10,-0.4) {$10$};
	\draw [dotted,color=black,thick] plot [smooth] coordinates {(0.1,0.1)(1,9)(9,9)(9,1)};
	\draw [dashed,color=black,thick] plot [smooth] coordinates {(0.1,0.1)(1,9)(9,1)(9,9)};
\end{tikzpicture}
\caption{Robot and specification workspace}
\label{fig:workspace}
\end{figure} 

We consider a single robot with double integrator dynamics on a planar plane as in Fig. \ref{fig:workspace}. The data for the system \eqref{system_discrete} with a sampling period of $0.5$ seconds is
$A=\begin{bmatrix}
1 & 0.5 &0 &0\\
0 &1& 0& 0\\
0& 0 &1 &0.5\\
0 &0& 0& 1
\end{bmatrix}$, $B=\begin{bmatrix}
0.125& 0 \\
0.5& 0 \\
0& 0.125 \\
0& 0.5
\end{bmatrix}$ and $\boldsymbol{x}(k)=\begin{bmatrix}
x& v_x& y &v_y
\end{bmatrix}^T$ denotes $x$-position, velocity in $x$-direction, $y$-position and velocity in $y$-direction, respectively. The input is constrained to $u \in [-1,1]\times [-1,1]$. The noise level can be characterized by the Signal-to-Noise ratio defined as
$
\text{SNR}_{\text{dB}} = 10 \cdot \operatorname{log}_{10}\Bigl(\frac{P_{\boldsymbol{x}}}{P_{\boldsymbol{v}}}\Bigr),
$
with $P_{\boldsymbol{x}}$ denoting the average signal power of $\boldsymbol{x}$. 

The specification imposed on the robot is to visit all three regions $A_1$, $A_2$ and $A_3$ (see Fig. \ref{fig:workspace}) within the time interval of $5$ to $25$ seconds while avoiding to leave the workspace as defined in $\psi_{i_4}$ below. The latter can be seen as a safety requirement. To use suitable predicates, consider the $p$-norm as $\|\boldsymbol{x}-\boldsymbol{x}_d\|_p<c$. The predicates (\ref{eq:stacked_pred}) are linear and therefore we can only deploy the infinity norm ($p=\infty$). Hence, the workspace can be separated into rectangles. Two possible paths $r_1$ and $r_2$ starting from $\boldsymbol{x}(0)=\begin{bmatrix}
0.1 & 0 & 0.1 & 0
\end{bmatrix}^T$ are depicted in Fig. \ref{fig:workspace}. As mentioned before, for robot motion planning formulas of the form $event \implies \psi_i$ are used. \ree{Simulation results for all-time satisfying formulas can be found in our previous work \cite{lindemann_1}}. Hence, the specification looks like
$
\phi = event \implies (\psi_{i_1} \wedge \psi_{i_2} \wedge \psi_{i_3} \wedge \psi_{i_4}),
$
where
$\psi_{i_1} = F_{[5,25]} (x \ge0 \wedge x\le2 \wedge y\ge8 \wedge y\le10)$, 
$\psi_{i_2} = F_{[5,25]} (x\ge8 \wedge x\le10 \wedge y\ge8 \wedge y\le10)$, 
$\psi_{i_3} = F_{[5,25]} (x\ge8 \wedge x\le10 \wedge y\ge0 \wedge y\le2)$ and
$\psi_{i_4} = G_{[0,25]} (x\ge0 \wedge x\le10 \wedge y\ge0 \wedge y\le10)$.

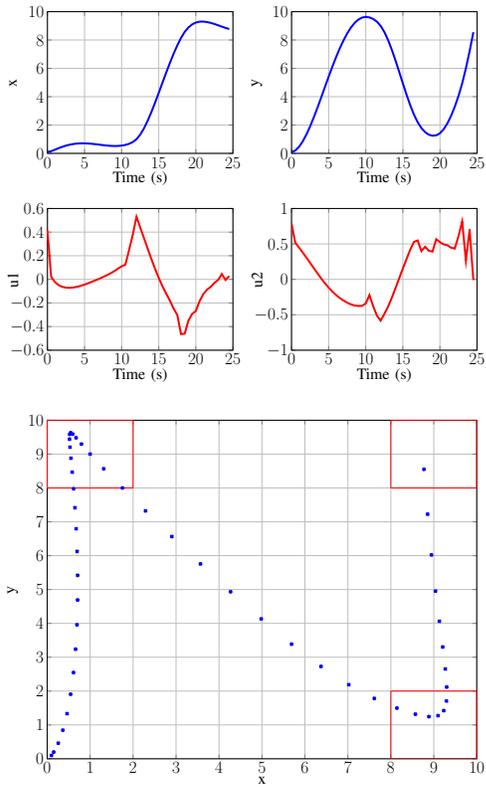
\begin{figure}
\centering
\subfloat{
%
%
\begin{tikzpicture}[scale=0.373]

\begin{axis}[%
width=2.603in,
height=1.99in,
at={(1.011in,3.406in)},
scale only axis,
xmin=0,
xmax=25,
xlabel={Time (s)},
xmajorgrids,
ymin=0,
ymax=10,
ylabel={x},
ymajorgrids,
axis background/.style={fill=white},
title style={font=\LARGE},xlabel style={font=\LARGE},ylabel style={font=\LARGE},legend style={font=\LARGE},ticklabel style={font=\LARGE},
]
\addplot [color=blue,solid,line width=2.0pt,forget plot]
  table[row sep=crcr]{%
0	0.1\\
0.5	0.151783776167884\\
1	0.258257249285716\\
1.5	0.364779997998159\\
2	0.462329708061167\\
2.5	0.545833879220368\\
3	0.612619223465221\\
3.5	0.66164125417093\\
4	0.693076597185683\\
4.5	0.708040600475645\\
5	0.708337429136591\\
5.5	0.696292052714056\\
6	0.674624683151137\\
6.5	0.646329898392776\\
7	0.614532907606594\\
7.5	0.582470991442088\\
8	0.553575542793331\\
8.5	0.531489202165095\\
9	0.520091028493881\\
9.5	0.523646385745476\\
10	0.547068782405963\\
10.5	0.595561873836967\\
11	0.673396137032208\\
11.5	0.797631682339853\\
12	0.999999999996272\\
12.5	1.3159080314723\\
13	1.7530871419362\\
13.5	2.28921461814668\\
14	2.9020142928588\\
14.5	3.56922552050517\\
15	4.26935442695839\\
15.5	4.98224268005089\\
16	5.68941606372549\\
16.5	6.37422954154669\\
17	7.02149360393597\\
17.5	7.61547597254597\\
18	8.14097758192901\\
18.5	8.5708882799957\\
19	8.88527585915326\\
19.5	9.09837298528857\\
20	9.23118150851333\\
20.5	9.29385496348339\\
21	9.29984621151459\\
21.5	9.267064640647\\
22	9.2076561303636\\
22.5	9.12930291377861\\
23	9.03925763722163\\
23.5	8.94477369828744\\
24	8.85524731253031\\
24.5	8.7706958618367\\
};
\end{axis}

\begin{axis}[%
width=2.603in,
height=1.99in,
at={(1.011in,0.642in)},
scale only axis,
xmin=0,
xmax=25,
xlabel={Time (s)},
xmajorgrids,
ymin=-0.6,
ymax=0.6,
ylabel={u1},
ymajorgrids,
axis background/.style={fill=white},
title style={font=\LARGE},xlabel style={font=\LARGE},ylabel style={font=\LARGE},legend style={font=\LARGE},ticklabel style={font=\LARGE},
]
\addplot [color=red,solid,line width=2.0pt,forget plot]
  table[row sep=crcr]{%
0	0.414270209343072\\
0.5	0.0232473662565091\\
1	-0.0228531614996178\\
1.5	-0.0489311476958622\\
2	-0.0634331635345902\\
2.5	-0.0703174517801981\\
3	-0.0717890565329579\\
3.5	-0.0689044449946825\\
4	-0.0628662728036481\\
4.5	-0.0544711242284844\\
5	-0.0442665164393548\\
5.5	-0.032709428683722\\
6	-0.0203098928798143\\
6.5	-0.00770775534275914\\
7	0.00558835231617505\\
7.5	0.0197433878098123\\
8	0.0347294763543631\\
8.5	0.0507758593018006\\
9	0.0688523880806797\\
9.5	0.0900839271904585\\
10	0.11048163097367\\
10.5	0.124247743140231\\
11	0.246962513759\\
11.5	0.378099665031193\\
12	0.53021804552567\\
12.5	0.439950586377293\\
13	0.351636339595411\\
13.5	0.261741248417664\\
14	0.173551175056329\\
14.5	0.0897902553984482\\
15	0.0122845177158308\\
15.5	-0.0580034730590041\\
16	-0.12087577376826\\
16.5	-0.17951954968703\\
17	-0.246734000547288\\
17.5	-0.301112073268297\\
18	-0.463615217262615\\
18.5	-0.46056973401032\\
19	-0.349753890167783\\
19.5	-0.292554933116548\\
20	-0.268525612921044\\
20.5	-0.184932042589905\\
21	-0.125250508600359\\
21.5	-0.0877650067261242\\
22	-0.0637926436865776\\
22.5	-0.0297438360893479\\
23	-0.00576546292836577\\
23.5	0.0454258883448573\\
24	-0.00562640783667978\\
24.5	0.0273212679861539\\
};
\end{axis}

\begin{axis}[%
width=2.603in,
height=1.99in,
at={(4.436in,3.406in)},
scale only axis,
xmin=0,
xmax=25,
xlabel={Time (s)},
xmajorgrids,
ymin=0,
ymax=10,
ylabel={y},
ymajorgrids,
axis background/.style={fill=white},
title style={font=\LARGE},xlabel style={font=\LARGE},ylabel style={font=\LARGE},legend style={font=\LARGE},ticklabel style={font=\LARGE},
]
\addplot [color=blue,solid,line width=2.0pt,forget plot]
  table[row sep=crcr]{%
0	0.1\\
0.5	0.198718793509797\\
1	0.461229185963232\\
1.5	0.845034807581166\\
2	1.33253007591787\\
2.5	1.90598199962394\\
3	2.5471649585495\\
3.5	3.23743553333013\\
4	3.95812190845991\\
4.5	4.69098425736601\\
5	5.41862132672765\\
5.5	6.12485209270627\\
6	6.7949866364622\\
6.5	7.41597098684199\\
7	7.97643431569599\\
7.5	8.4666081123866\\
8	8.87822309474431\\
8.5	9.20454895581719\\
9	9.44059729514806\\
9.5	9.58350271405126\\
10	9.63263257578141\\
10.5	9.59242233717863\\
11	9.48189646280149\\
11.5	9.29640715039837\\
12	9.00000000000365\\
12.5	8.56693925414284\\
13	7.99960076746251\\
13.5	7.32241456622683\\
14	6.56366971265978\\
14.5	5.75593998472981\\
15	4.93394538453888\\
15.5	4.1326481913944\\
16	3.38607318328298\\
16.5	2.72769633325989\\
17	2.18794011551054\\
17.5	1.78292140707946\\
18	1.49634378987327\\
18.5	1.3171250658194\\
19	1.24587550705778\\
19.5	1.27407320483026\\
20	1.4221630682603\\
20.5	1.70681487269711\\
21	2.11886178880498\\
21.5	2.65276870618127\\
22	3.30263449769984\\
22.5	4.06255309246116\\
23	4.9528691761278\\
23.5	6.02140959812497\\
24	7.22529161416282\\
24.5	8.55040882963837\\
};
\end{axis}

\begin{axis}[%
width=2.603in,
height=1.99in,
at={(4.436in,0.642in)},
scale only axis,
xmin=0,
xmax=25,
xlabel={Time (s)},
xmajorgrids,
ymin=-1,
ymax=1,
ylabel={u2},
ymajorgrids,
axis background/.style={fill=white},
title style={font=\LARGE},xlabel style={font=\LARGE},ylabel style={font=\LARGE},legend style={font=\LARGE},ticklabel style={font=\LARGE},
]
\addplot [color=red,solid,line width=2.0pt,forget plot]
  table[row sep=crcr]{%
0	0.789750348078378\\
0.5	0.520582443470718\\
1	0.449779389845286\\
1.5	0.379737783904827\\
2	0.307915459050201\\
2.5	0.233932822705655\\
3	0.158768104134921\\
3.5	0.0845582986582301\\
4	0.0128494915523747\\
4.5	-0.0546517279080853\\
5	-0.116598699156017\\
5.5	-0.172171078625558\\
6	-0.221030468383493\\
6.5	-0.263137703822892\\
7	-0.29917855348419\\
7.5	-0.329291961179077\\
8	-0.353021009099483\\
8.5	-0.369199164836641\\
9	-0.375944198584721\\
9.5	-0.374260258799638\\
10	-0.340460543863913\\
10.5	-0.222064542330866\\
11	-0.377642961876931\\
11.5	-0.509699742056017\\
12	-0.583529021672591\\
12.5	-0.490692904883568\\
13	-0.388088811559291\\
13.5	-0.264380407091635\\
14	-0.127498587811749\\
14.5	0.0133796097241359\\
15	0.152199646647417\\
15.5	0.285577833617018\\
16	0.420007431089663\\
16.5	0.528957627100268\\
17	0.548942447445925\\
17.5	0.398586282353174\\
18	0.46028486286535\\
18.5	0.403468459472639\\
19	0.392109592800213\\
19.5	0.567027732460258\\
20	0.525467795593836\\
20.5	0.493693097774755\\
21	0.481186912372529\\
21.5	0.446484080765711\\
22	0.433938345176293\\
22.5	0.609241566066338\\
23	0.816553140577774\\
23.5	0.266179611747803\\
24	0.703701983753704\\
24.5	-0.0082908818858326\\
};
\end{axis}
\end{tikzpicture}%
} \\
\subfloat{
%
%
\begin{tikzpicture}[scale=0.373]

\begin{axis}[%
width=6.028in,
height=4.754in,
at={(1.011in,0.642in)},
scale only axis,
xmin=0,
xmax=10,
xlabel={x},
xmajorgrids,
ymin=0,
ymax=10,
ylabel={y},
ymajorgrids,
axis background/.style={fill=white},
title style={font=\LARGE},xlabel style={font=\LARGE},ylabel style={font=\LARGE},legend style={font=\LARGE},ticklabel style={font=\LARGE},
]
\addplot [color=blue,line width=2.0pt,only marks,mark=x,mark options={solid},forget plot]
  table[row sep=crcr]{%
0.1	0.1\\
0.151783776167884	0.198718793509797\\
0.258257249285716	0.461229185963232\\
0.364779997998159	0.845034807581166\\
0.462329708061167	1.33253007591787\\
0.545833879220368	1.90598199962394\\
0.612619223465221	2.5471649585495\\
0.66164125417093	3.23743553333013\\
0.693076597185683	3.95812190845991\\
0.708040600475645	4.69098425736601\\
0.708337429136591	5.41862132672765\\
0.696292052714056	6.12485209270627\\
0.674624683151137	6.7949866364622\\
0.646329898392776	7.41597098684199\\
0.614532907606594	7.97643431569599\\
0.582470991442088	8.4666081123866\\
0.553575542793331	8.87822309474431\\
0.531489202165095	9.20454895581719\\
0.520091028493881	9.44059729514806\\
0.523646385745476	9.58350271405126\\
0.547068782405963	9.63263257578141\\
0.595561873836967	9.59242233717863\\
0.673396137032208	9.48189646280149\\
0.797631682339853	9.29640715039837\\
0.999999999996272	9.00000000000365\\
1.3159080314723	8.56693925414284\\
1.7530871419362	7.99960076746251\\
2.28921461814668	7.32241456622683\\
2.9020142928588	6.56366971265978\\
3.56922552050517	5.75593998472981\\
4.26935442695839	4.93394538453888\\
4.98224268005089	4.1326481913944\\
5.68941606372549	3.38607318328298\\
6.37422954154669	2.72769633325989\\
7.02149360393597	2.18794011551054\\
7.61547597254597	1.78292140707946\\
8.14097758192901	1.49634378987327\\
8.5708882799957	1.3171250658194\\
8.88527585915326	1.24587550705778\\
9.09837298528857	1.27407320483026\\
9.23118150851333	1.4221630682603\\
9.29385496348339	1.70681487269711\\
9.29984621151459	2.11886178880498\\
9.267064640647	2.65276870618127\\
9.2076561303636	3.30263449769984\\
9.12930291377861	4.06255309246116\\
9.03925763722163	4.9528691761278\\
8.94477369828744	6.02140959812497\\
8.85524731253031	7.22529161416282\\
8.7706958618367	8.55040882963837\\
};
\addplot [color=red,solid,forget plot]
  table[row sep=crcr]{%
0	8\\
2	8\\
};
\addplot [color=red,solid,forget plot]
  table[row sep=crcr]{%
0	10\\
2	10\\
};
\addplot [color=red,solid,forget plot]
  table[row sep=crcr]{%
0	8\\
0	10\\
};
\addplot [color=red,solid,forget plot]
  table[row sep=crcr]{%
2	8\\
2	10\\
};
\addplot [color=red,solid,forget plot]
  table[row sep=crcr]{%
8	8\\
10	8\\
};
\addplot [color=red,solid,forget plot]
  table[row sep=crcr]{%
8	10\\
10	10\\
};
\addplot [color=red,solid,forget plot]
  table[row sep=crcr]{%
8	8\\
8	10\\
};
\addplot [color=red,solid,forget plot]
  table[row sep=crcr]{%
10	8\\
10	10\\
};
\addplot [color=red,solid,forget plot]
  table[row sep=crcr]{%
8	0\\
10	0\\
};
\addplot [color=red,solid,forget plot]
  table[row sep=crcr]{%
8	2\\
10	2\\
};
\addplot [color=red,solid,forget plot]
  table[row sep=crcr]{%
8	0\\
8	2\\
};
\addplot [color=red,solid,forget plot]
  table[row sep=crcr]{%
10	0\\
10	2\\
};
\end{axis}
\end{tikzpicture}%
}
\caption{$\text{SNR}_{\text{dB}}=\infty$ dB}
\label{fig:1}
\end{figure}

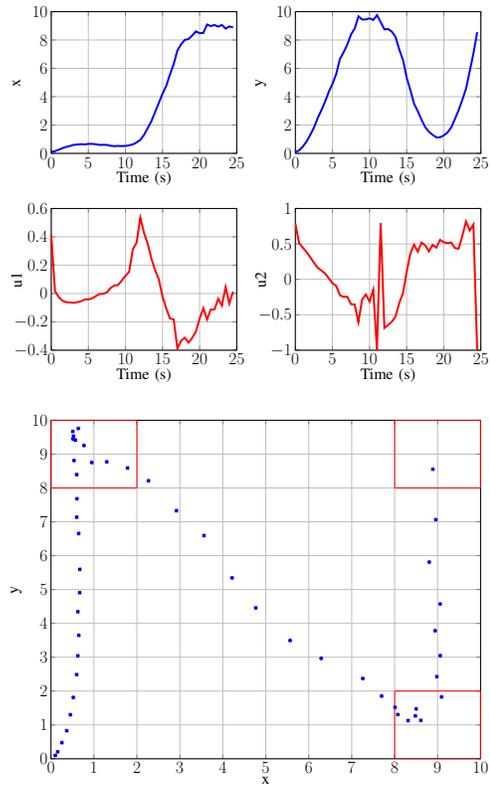
\begin{figure}
\centering
\subfloat{
%
%
\begin{tikzpicture}[scale=0.373]

\begin{axis}[%
width=2.603in,
height=1.99in,
at={(1.011in,3.406in)},
scale only axis,
xmin=0,
xmax=25,
xlabel={Time (s)},
xmajorgrids,
ymin=0,
ymax=10,
ylabel={x},
ymajorgrids,
axis background/.style={fill=white},
title style={font=\LARGE},xlabel style={font=\LARGE},ylabel style={font=\LARGE},legend style={font=\LARGE},ticklabel style={font=\LARGE},
]
\addplot [color=blue,solid,line width=2.0pt,forget plot]
  table[row sep=crcr]{%
0	0.1\\
0.5	0.156292321275757\\
1	0.254429147132502\\
1.5	0.366245231572847\\
2	0.452084256806895\\
2.5	0.519138057464854\\
3	0.598064625910342\\
3.5	0.625365393934935\\
4	0.647505155531402\\
4.5	0.625656846485234\\
5	0.668263151066774\\
5.5	0.671312318859528\\
6	0.645673757610382\\
6.5	0.598899589361798\\
7	0.604750253306858\\
7.5	0.599190579536918\\
8	0.537163594838128\\
8.5	0.509017428767817\\
9	0.526634662382136\\
9.5	0.51444541433895\\
10	0.52331226813936\\
10.5	0.568993563831836\\
11	0.636119584594236\\
11.5	0.768550249553715\\
12	0.949836478077178\\
12.5	1.29778991577311\\
13	1.78189318551004\\
13.5	2.26880529579556\\
14	2.91977861118623\\
14.5	3.56138987148058\\
15	4.21535560389941\\
15.5	4.76729635302342\\
16	5.56682366007757\\
16.5	6.28942496127958\\
17	7.25797358837429\\
17.5	7.69600989835919\\
18	8.00832946170334\\
18.5	8.07379463326872\\
19	8.31264844042337\\
19.5	8.60958231053141\\
20	8.48160485266011\\
20.5	8.50031075325313\\
21	9.09108378343025\\
21.5	8.9830116681979\\
22	9.06062438358378\\
22.5	8.94360837345348\\
23	9.05953319743517\\
23.5	8.80586205993599\\
24	8.95758298087125\\
24.5	8.88951277474556\\
};
\end{axis}

\begin{axis}[%
width=2.603in,
height=1.99in,
at={(1.011in,0.642in)},
scale only axis,
xmin=0,
xmax=25,
xlabel={Time (s)},
xmajorgrids,
ymin=-0.4,
ymax=0.6,
ylabel={u1},
ymajorgrids,
axis background/.style={fill=white},
title style={font=\LARGE},xlabel style={font=\LARGE},ylabel style={font=\LARGE},legend style={font=\LARGE},ticklabel style={font=\LARGE},
]
\addplot [color=red,solid,line width=2.0pt,forget plot]
  table[row sep=crcr]{%
0	0.414270209343072\\
0.5	0.0180219244251507\\
1	-0.0249132951103421\\
1.5	-0.050906541562747\\
2	-0.060933572124299\\
2.5	-0.0629226662142817\\
3	-0.0658555031042311\\
3.5	-0.061621908427117\\
4	-0.0547966124784209\\
4.5	-0.0417822532656563\\
5	-0.0418252368119447\\
5.5	-0.034586572159173\\
6	-0.0186042420479422\\
6.5	-0.00362265502616632\\
7	-0.00188637996646612\\
7.5	0.00820356532901645\\
8	0.0394164498093865\\
8.5	0.0565160232898161\\
9	0.0598753701278107\\
9.5	0.0910485084913082\\
10	0.124039468374349\\
10.5	0.153732917701232\\
11	0.313039467786662\\
11.5	0.358671629014864\\
12	0.534262793522906\\
12.5	0.428951833642104\\
13	0.351093311665843\\
13.5	0.246171595971984\\
14	0.164395800334859\\
14.5	0.0993724947742434\\
15	-0.0222962100328089\\
15.5	-0.10565417408122\\
16	-0.176541318023197\\
16.5	-0.182750668294464\\
17	-0.388599350740836\\
17.5	-0.333341340315178\\
18	-0.31188349729079\\
18.5	-0.346958253411887\\
19	-0.312631069484351\\
19.5	-0.268137378326322\\
20	-0.176695845190385\\
20.5	-0.102717753574165\\
21	-0.182133181656732\\
21.5	-0.112979741152926\\
22	-0.112381099490887\\
22.5	-0.0370576051920146\\
23	-0.0845063970037432\\
23.5	0.0457453464190112\\
24	-0.0717334578626182\\
24.5	0.0134890369930432\\
};
\end{axis}

\begin{axis}[%
width=2.603in,
height=1.99in,
at={(4.436in,3.406in)},
scale only axis,
xmin=0,
xmax=25,
xlabel={Time (s)},
xmajorgrids,
ymin=0,
ymax=10,
ylabel={y},
ymajorgrids,
axis background/.style={fill=white},
title style={font=\LARGE},xlabel style={font=\LARGE},ylabel style={font=\LARGE},legend style={font=\LARGE},ticklabel style={font=\LARGE},
]
\addplot [color=blue,solid,line width=2.0pt,forget plot]
  table[row sep=crcr]{%
0	0.1\\
0.5	0.206395249331295\\
1	0.476701213222181\\
1.5	0.828976467466653\\
2	1.30174233149433\\
2.5	1.81113380670203\\
3	2.48518944847226\\
3.5	3.0416178541921\\
4	3.64605702641871\\
4.5	4.34390002889134\\
5	4.90726049082528\\
5.5	5.59498048117649\\
6	6.65446205295952\\
6.5	7.13599264022344\\
7	7.68106880823568\\
7.5	8.39441763228709\\
8	8.81036406542744\\
8.5	9.67099663583886\\
9	9.44500787528176\\
9.5	9.45141894267565\\
10	9.53216704871219\\
10.5	9.41214070564193\\
11	9.75726290483261\\
11.5	9.25370516382392\\
12	8.75139729292409\\
12.5	8.77228014075019\\
13	8.58899823310674\\
13.5	8.21258066344091\\
14	7.32839216549028\\
14.5	6.59380965851094\\
15	5.34331774390555\\
15.5	4.45737907453538\\
16	3.4949553975403\\
16.5	2.96591687864247\\
17	2.37116365615032\\
17.5	1.85258116467346\\
18	1.52065774706152\\
18.5	1.30668482450256\\
19	1.12911893023582\\
19.5	1.13435423585394\\
20	1.2677406399259\\
20.5	1.47284433162766\\
21	1.82704318884255\\
21.5	2.42594739822248\\
22	3.04595082429315\\
22.5	3.7826837274599\\
23	4.57068054974352\\
23.5	5.80981267407513\\
24	7.0639388824367\\
24.5	8.55270329476711\\
};
\end{axis}

\begin{axis}[%
width=2.603in,
height=1.99in,
at={(4.436in,0.642in)},
scale only axis,
xmin=0,
xmax=25,
xlabel={Time (s)},
xmajorgrids,
ymin=-1,
ymax=1,
ylabel={u2},
ymajorgrids,
axis background/.style={fill=white},
title style={font=\LARGE},xlabel style={font=\LARGE},ylabel style={font=\LARGE},legend style={font=\LARGE},ticklabel style={font=\LARGE},
]
\addplot [color=red,solid,line width=2.0pt,forget plot]
  table[row sep=crcr]{%
0	0.789750348078378\\
0.5	0.515478942931278\\
1	0.448174190571251\\
1.5	0.386904826264645\\
2	0.313530786475007\\
2.5	0.235690201656692\\
3	0.164917354669393\\
3.5	0.125326238731948\\
4	0.0839764779522294\\
4.5	0.0155818518808111\\
5	-0.0549272976407168\\
5.5	-0.0903626034336559\\
6	-0.223519548854672\\
6.5	-0.248693015577401\\
7	-0.244561749445167\\
7.5	-0.350836160253308\\
8	-0.354699446782225\\
8.5	-0.603439756027982\\
9	-0.284508005363525\\
9.5	-0.212426788946258\\
10	-0.314521781754166\\
10.5	-0.145273550927449\\
11	-0.999999999171228\\
11.5	0.795552033532133\\
12	-0.687229943621385\\
12.5	-0.644889767244587\\
13	-0.598610543549187\\
13.5	-0.52566662876672\\
14	-0.334750172343209\\
14.5	-0.195931643318619\\
15	0.0998113060467163\\
15.5	0.36521450011085\\
16	0.489135301461928\\
16.5	0.391006554254168\\
17	0.518592185190059\\
17.5	0.480327245750014\\
18	0.395024554225889\\
18.5	0.486789791372467\\
19	0.448067011484911\\
19.5	0.55711766610409\\
20	0.524460914937249\\
20.5	0.509960180971846\\
21	0.520503481783698\\
21.5	0.442693297055275\\
22	0.430815009396934\\
22.5	0.609922811065873\\
23	0.816620095306826\\
23.5	0.688908017070601\\
24	0.774637325830653\\
24.5	-0.998436156541498\\
};
\end{axis}
\end{tikzpicture}%
} \\
\subfloat{
%
%
\begin{tikzpicture}[scale=0.373]

\begin{axis}[%
width=6.028in,
height=4.754in,
at={(1.011in,0.642in)},
scale only axis,
xmin=0,
xmax=10,
xlabel={x},
xmajorgrids,
ymin=0,
ymax=10,
ylabel={y},
ymajorgrids,
axis background/.style={fill=white},
title style={font=\LARGE},xlabel style={font=\LARGE},ylabel style={font=\LARGE},legend style={font=\LARGE},ticklabel style={font=\LARGE},
]
\addplot [color=blue,line width=2.0pt,only marks,mark=x,mark options={solid},forget plot]
  table[row sep=crcr]{%
0.1	0.1\\
0.156292321275757	0.206395249331295\\
0.254429147132502	0.476701213222181\\
0.366245231572847	0.828976467466653\\
0.452084256806895	1.30174233149433\\
0.519138057464854	1.81113380670203\\
0.598064625910342	2.48518944847226\\
0.625365393934935	3.0416178541921\\
0.647505155531402	3.64605702641871\\
0.625656846485234	4.34390002889134\\
0.668263151066774	4.90726049082528\\
0.671312318859528	5.59498048117649\\
0.645673757610382	6.65446205295952\\
0.598899589361798	7.13599264022344\\
0.604750253306858	7.68106880823568\\
0.599190579536918	8.39441763228709\\
0.537163594838128	8.81036406542744\\
0.509017428767817	9.67099663583886\\
0.526634662382136	9.44500787528176\\
0.51444541433895	9.45141894267565\\
0.52331226813936	9.53216704871219\\
0.568993563831836	9.41214070564193\\
0.636119584594236	9.75726290483261\\
0.768550249553715	9.25370516382392\\
0.949836478077178	8.75139729292409\\
1.29778991577311	8.77228014075019\\
1.78189318551004	8.58899823310674\\
2.26880529579556	8.21258066344091\\
2.91977861118623	7.32839216549028\\
3.56138987148058	6.59380965851094\\
4.21535560389941	5.34331774390555\\
4.76729635302342	4.45737907453538\\
5.56682366007757	3.4949553975403\\
6.28942496127958	2.96591687864247\\
7.25797358837429	2.37116365615032\\
7.69600989835919	1.85258116467346\\
8.00832946170334	1.52065774706152\\
8.07379463326872	1.30668482450256\\
8.31264844042337	1.12911893023582\\
8.60958231053141	1.13435423585394\\
8.48160485266011	1.2677406399259\\
8.50031075325313	1.47284433162766\\
9.09108378343025	1.82704318884255\\
8.9830116681979	2.42594739822248\\
9.06062438358378	3.04595082429315\\
8.94360837345348	3.7826837274599\\
9.05953319743517	4.57068054974352\\
8.80586205993599	5.80981267407513\\
8.95758298087125	7.0639388824367\\
8.88951277474556	8.55270329476711\\
};
\addplot [color=red,solid,forget plot]
  table[row sep=crcr]{%
0	8\\
2	8\\
};
\addplot [color=red,solid,forget plot]
  table[row sep=crcr]{%
0	10\\
2	10\\
};
\addplot [color=red,solid,forget plot]
  table[row sep=crcr]{%
0	8\\
0	10\\
};
\addplot [color=red,solid,forget plot]
  table[row sep=crcr]{%
2	8\\
2	10\\
};
\addplot [color=red,solid,forget plot]
  table[row sep=crcr]{%
8	8\\
10	8\\
};
\addplot [color=red,solid,forget plot]
  table[row sep=crcr]{%
8	10\\
10	10\\
};
\addplot [color=red,solid,forget plot]
  table[row sep=crcr]{%
8	8\\
8	10\\
};
\addplot [color=red,solid,forget plot]
  table[row sep=crcr]{%
10	8\\
10	10\\
};
\addplot [color=red,solid,forget plot]
  table[row sep=crcr]{%
8	0\\
10	0\\
};
\addplot [color=red,solid,forget plot]
  table[row sep=crcr]{%
8	2\\
10	2\\
};
\addplot [color=red,solid,forget plot]
  table[row sep=crcr]{%
8	0\\
8	2\\
};
\addplot [color=red,solid,forget plot]
  table[row sep=crcr]{%
10	0\\
10	2\\
};
\end{axis}
\end{tikzpicture}%
}
\caption{$\text{SNR}_{\text{dB}}=16.23 $ dB}
\label{fig:2}
\end{figure}

Fig. \ref{fig:1} shows the MPC result in case of no noise, whereas Fig. \ref{fig:2} depicts the result for the disturbed case. In both figures, the upper subfigure shows the $x$ and $y$ evolution and the corresponding inputs separately, whereas the lower subfigure shows the resulting trajectory. The SNR for this example is $16.23$ dB and satisfaction is still ensured due to the robust MPC implementation. The proposed MPC provides optimal robustness in the sense that it steers the state trajectory in the direction, where it has the farthest distance to the set of states not fulfilling the formula. \re{Note the computational ease compared with the non-convex MILP implementation of \cite{raman1} and \cite{sadraddini} where it is possible to maximize Space Robustness.}

\section{Discussion and Future Work}
\label{sec:diskussion_futurework}
\re{This paper introduced Discrete Average Space Robustness as new robust semantics for Signal Temporal Logic. These semantics are linear in the temporal operators and hence easier to use in control synthesis}. Robust Control is achieved due to direct maximization of Discrete Average Space Robustness. Motion planning has been considered by using the infinity norm and one-time satisfying formulas. This was depicted in simulations for a single agent with double integrator dynamics in a planar workspace. 

In the current framework, we have not considered obstacles within the workspace. This is subject to future work and could potentially be handled in different ways. Future work will also include an extension to multi-agent systems. An advantage compared with the traditional point-to-point navigation objective is that this methodology can include other specifications in a rather straightforward manner. For instance, multiple destinations can be visited (periodically), while robot specific requirements can easily be added. Furthermore, the proposed methodology has low computation times due to the Linear Program, considers average performance and results in a robustness against model uncertainties and noise.

%
%



\bibliographystyle{IEEEtran}
\bibliography{literature}

\end{document}